# The management of obstructive sleep apnea accompanied with mandibular retrognathia across the lifespan from orthodontic perspective

Meng-Han Zhang[1,2], Yue-Hua Liu[2*]

[1]School of Stomatology affiliated to Medical College, Zhejiang University, Hangzhou, China

[2]Shanghai Key Laboratory of Craniomaxillofacial Development and Diseases, Shanghai Stomatological Hospital & School of Stomatology, Fudan University, Shanghai, China

* Correspondence: Yue-Hua Liu (liuyuehua@fudan.edu.cn)

Yue-Hua Liu, Department of Orthodontics, Shanghai Stomatological Hospital, Fudan University, Shanghai 200001, China. Tel: +86-21-55664116; E-mail: liuyuehua@fudan.edu.cn.

**Abstract**

Obstructive sleep apnea (OSA) is a complex disease with complex etiology, which requires multidisciplinary cooperation in diagnosis and treatment. Mandibular retrognathia is strongly associated with OSA. Orthodontists can either correct the mandibular retrognathia of pediatric OSA via various kinds of orthodontic appliances, following adenoidectomy and tonsillectomy, or enlarge upper airway by mandibular advancement device (MAD) through repositioning the mandible and tongue of adult OSA patients. This mini review was to investigate the therapy of MAD to adult OSA as well as orthodontic treatment to pediatric OSA.

Obstructive sleep apnea syndrome (OSA) is a sleep disorder with complex etiology, characterized by the disruption of the upper airway during sleep, resulting in sleep apnea, desaturation of oxyhemoglobin, and daytime sleepiness[1-2]. Chronic intermittent hypoxemia and sleep disorders result from affect the quality of life of patients and increase the risk of cardiovascular disease, neurocognitive dysfunction, metabolic disorders and sudden death[3-5].

Dentofacial deformities are considered an important risk factor in OAS patients. Mandibular retrognathia, one of the most common dentofacial deformities, is responsible for the reduction in posterior airway space causing hypopharynx obstruction[6]. On the hand, OSA patients often present with mandibular retrognathia[7-8]. Orthodontists can either correct the dentofacial malocclusion (mandibular retrognathia) of children or juveniles via various kinds of appliances, following adenoidectomy and tonsillectomy, or enlarge upper airway through repositioning the mandible and tongue of an adult patient[9]. The research group is committed to studying OSA accompanied with mandibular retrognathia, and attempts to implement upper airway management across the lifespan from orthodontic perspective.

1. **Mandibular advancement device forward positioning mandible for adult OSA accompanied with mandibular retrognathia**

For adult OSA patients with mandibular retrognathia, continuous positive airway pressure (CPAP), and mandibular advancement device (MAD) are both widely used. A meta-analysis was conducted and revealed that patients opted for MAD compared with CPAP, especially adjustable MAD[10]. MAD can move the mandible and tongue forward, thus anatomically enlarge the oropharyngeal airway, which can compensate for upper airway fatigue. Our previews studies found adjustable MAD (Klearway™ appliance) could improve AHI by its mechanical expansion of the upper airway and stabilization of the jaw posture using supine cephalometric measurements[11]. The research group also test the effect of custom-made device with different mandible advancement distance by assessing MRI, and found increase in its transverse dimension of the velopharynx in awake OSA patients[12]. The research group reviewed and compared various kinds of oral appliances[13-14], then developed an adjustable MAD (Patent No. CN2418850Y).

However, the upper airway obstruction is dynamic and multilevel in OSA patients during sleep according to dynamic MRI reports[15]. Therefore, we developed a computer-aided mandibular repositioning system (CAMRS) (Patent No. CN1602970A, No. CN100486545C), consist of a polysomnography monitor, an MAD, a transmission device and a control computer. CAMRS could reproductively and effectively reposition mandible with little sleeping disturbance[16]. Nowadays, we developed an intraoral invisible drug slow-release/ventilation device for improving apnea syndrome, which is comprises a tooth socket unit and the functional unit with sustained release of medicaments (Patent No. CN111135438A). We hope to apply MAD as an auxiliary drug delivery device.

2. **Orthodontic appliance accompanied with orofacial myofunctional therapy as an effective treatment for pediatric OSA with mandibular retrognathia**

Most pediatric OSA can be attribute to hypertrophy of the tonsils and/or adenoids, and surgical resection is one of the effective therapies for treating pediatric OSA[17]. Compared to the complexity and recurrence of treatment for adult OSA, the early prevention and intervention for pediatric OSA are simpler and more effective. OSA lead to compensatory oral breathing, and will induce a series of dentofacial anomalies. The earlier the intervention of orthodontic treatment, the simpler of therapy and the better prognosis would be. The research group performed a cross-sectional study of a total of 715 children, and found tonsil hypertrophy plays an essential role in the direction of dentofacial development[18]. Additionally, a total of 3433 subjects were studied and found the prevalence of OSA in primary students in the Chinese urban population was highly associated with mandibular retrognathia[19].

Therefore, the multidisciplinary combination treatment is appreciated by clinicians, such as orthodontic treatment and surgical treatment for OSA in children with mandible retrusion. The research group introduced various orthodontic treatment of OSA in children[20]. Maxillary expansion combined with Twin-block appliance can increase the upper airway volume of children with mouth breathing and malocclusion. The upper airway shape tends to be oval, respiratory symptoms are improved, which help the children and help establish normal nasal breathing[21]. Orofacial myofunctional therapy can correct adenoid face of children with OSA after adenotonsillectomy. The length of the upper lip is increased, the protrusion of the upper lip is reduced, and lips can close naturally. Orofacial myofunctional therapy can be considered as an effective complementary treatment for OSA patients with oral breathing after adenotonsillectomy[22]. Nowadays, a randomised, open-label, parallel-group, active controlled trial was designed to study the efficacy of orthodontic treatment versus adenotonsillectomy in children with mild or moderate OSA accompanied by tonsillar adenoid hypertrophy and mandibular retrognathia[23-24]. (ClinicalTrials.gov: NCT03451318).

3. **Summary**

The research group have been dedicated to the pathological mechanism and treatment of adult OSA and pediatric OSA for decades. Orthodontists can play an independent or collaborative therapeutic role in the management of the upper airway across the lifespan.

**Declaration of interests**

The authors declare no competing interests.